\documentclass[11pt]{article}

\usepackage{amsmath}
\usepackage{graphicx}
\usepackage{amsfonts}
\usepackage{amssymb}
\usepackage{epsfig}
\usepackage{color}
\usepackage{psfrag}
\usepackage{epstopdf}

\newcommand{\EQ}{\begin{equation}}
\newcommand{\EN}{\end{equation}}
\newcommand{\be}{\begin{equation}}
\newcommand{\ee}{\end{equation}}
\newcommand{\bea}{\begin{eqnarray}}
\newcommand{\eea}{\end{eqnarray}}

\begin{document} \setcounter{page}{0}
\topmargin 0pt
\oddsidemargin 5mm
\renewcommand{\thefootnote}{\arabic{footnote}}
\newpage
\setcounter{page}{0}
\topmargin 0pt
\oddsidemargin 5mm
\renewcommand{\thefootnote}{\arabic{footnote}}
\newpage
\begin{titlepage}
\begin{flushright}
\end{flushright}
\vspace{0.5cm}
\begin{center}
{\large {\bf Exact results for the $O(N)$ model with quenched disorder}}\\
\vspace{1.8cm}
{\large Gesualdo Delfino$^{1,2}$ and Noel Lamsen$^{1,2}$}\\
\vspace{0.5cm}
{\em $^1$SISSA -- Via Bonomea 265, 34136 Trieste, Italy}\\
{\em $^2$INFN sezione di Trieste}\\
\end{center}
\vspace{1.2cm}

\renewcommand{\thefootnote}{\arabic{footnote}}
\setcounter{footnote}{0}

\begin{abstract}
\noindent
We use scale invariant scattering theory to exactly determine the lines of renormalization group fixed points for $O(N)$-symmetric models with quenched disorder in two dimensions. Random fixed points are characterized by two disorder parameters: a modulus that vanishes when approaching the pure case, and a phase angle.  The critical lines fall into three classes depending on the values of the disorder modulus. Besides the class corresponding to the pure case, a second class has maximal value of the disorder modulus and includes Nishimori-like multicritical points as well as zero temperature fixed points. The third class contains critical lines that interpolate, as $N$ varies, between the first two classes. For positive $N$, it contains a single line of infrared fixed points spanning the values of $N$ from $\sqrt{2}-1$ to $1$. The symmetry sector of the energy density operator is superuniversal (i.e. $N$-independent) along this line. For $N=2$ a line of fixed points exists only in the pure case, but accounts also for the Berezinskii-Kosterlitz-Thouless phase observed in presence of disorder. 
\end{abstract}
\end{titlepage}

\newpage
Gaining theoretical access to the critical properties of disordered systems with short range interactions has been a challenging problem of statistical mechanics. For weak randomness, the Harris criterion \cite{Harris} relates the relevance of disorder to the sign of the specific heat critical exponent of the pure system. If this sign is positive weak disorder drives the system towards a new (``random'') fixed point of the renormalization group, responsible for new critical exponents that in some limits can be computed perturbatively (see e.g. \cite{Cardy_book}). In the regime of strong disorder, a relevant role is played by the gauge symmetry \cite{Nishimori} exhibited by systems such as the Ising model with $\pm J$ bond randomness. This allows, in particular, the idenfitication of a multicritical point along the phase boundary separating the ferromagnetic and the paramagnetic (or spin glass, if present) phases in the temperature-disorder plane. For the rest, the study of critical properties at strong disorder has essentially relied on numerical methods.

Particularly noticeable has been the absence of exact results in two dimensions, to the point that one could legitimately wonder whether random fixed points of planar systems possess the infinite-dimensional conformal symmetry \cite{BPZ,DfMS} that yielded the exact critical exponents in the pure case. Progress has been achieved recently \cite{random} extending to the random case the idea of implementing conformal invariance within the basis of the underlying particle excitations \cite{paraf,fpu}. It was explicitly shown in \cite{random,potts_random} for the $q$-state Potts model with quenched disorder how the method yields exact equations for the scattering amplitudes whose solutions correspond to random fixed points. One of the remarkable emerging properties is the presence of superuniversal (i.e. symmetry independent) sectors able to shed light on longstanding numerical and theoretical puzzles for critical exponents.

In this paper we consider two-dimensional disordered systems with $O(N)$ symmetry that reduce to the $N$-vector ferromagnet in the pure limit. It is known that weak disorder is marginally irrelevant at $N=1$ (Ising) \cite{DD}, and becomes relevant for $N<1$. This means that slightly below $N=1$ an infrared random fixed point can be found through a perturbative approach similar to that used in \cite{Ludwig,DPP} for the $q\to 2^+$ Potts model. This perturbative study was performed in \cite{Shimada}, where the one-loop beta function was used to argue that the line of infrared fixed points spans an interval $N\in(N_*,1)$, while in the interval $N\in(0,N_*)$ the system flows directly to a strong disorder regime; the estimate $N_*\approx 0.26$ was obtained within the one-loop approximation. The $O(N)$ model with a specific bimodal distribution of bond disorder was then studied in \cite{SJK} within a numerical transfer matrix approach. In particular, this study confirmed the presence of the lower endpoint $N_*$ for the line of infrared fixed points originating at $N=1$, and obtained the estimate $N_*\approx 0.5$. At $N=N_*$ the infrared line was observed to join a line of strong randomness multicritical points extending for $N>N_*$, and the universal properties of the point at $N=1$ on this line were found in quantitative agreement with those of the Nishimori multicritical point. 

Below we will use the scattering formalism to exactly determine the lines of renormalization group fixed points for systems with $O(N)$ symmetry in presence of quenched disorder. We will show, in particular, that these critical lines  belong to three different classes depending on the values of a disorder modulus $\rho_4$, one of two parameters associated to disorder. The three classes are: solutions for the pure systems ($\rho_4=0$), strongly disordered solutions ($\rho_4=1$), and solutions with values of $\rho_4$ interpolating between $0$ and $1$. For positve $N$, the latter class contains a single line of infrared fixed points, extending from $N=1$ (where $\rho_4=0$) down to $N=N_*=\sqrt{2}-1=0.414..$ (where $\rho_4=1$). At $N_*$ this line joins one of the solutions in the class $\rho_4=1$, which are defined for any $N$. The class with $\rho_4=1$ contains fixed points that do not merge a fixed point of the pure system in any limit. Typical examples in this class are the multicritical points of Nishimori type and those encountered flowing from such a multicritical point towards lower temperatures. 

\begin{figure}
\begin{center}
\includegraphics[width=5cm]{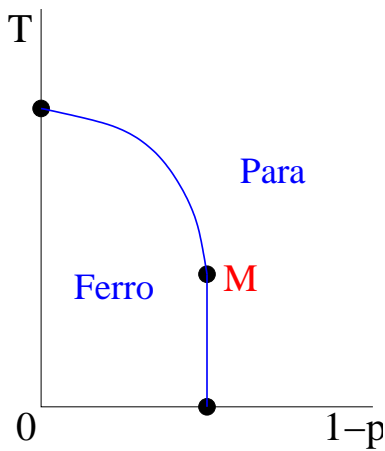}
\caption{Qualitative phase diagram and expected fixed points for the two-dimensional Ising model with $\pm J$ disorder. $1-p$ is the amount of disorder and $M$ indicates the multicritical (Nishimori) point. 
}
\label{phd}
\end{center} 
\end{figure}

We start recalling that the random bond $N$-vector model is defined by the lattice Hamiltonian
\EQ
{\cal H}=-\sum_{\langle i,j\rangle}J_{ij}\,{\bf s}_i\cdot{\bf s}_j\,,
\EN
where ${\bf s}_i$ is a $N$-component unit vector located at site $i$, the sum runs over nearest neighboring sites, and $J_{ij}$ are bond couplings drawn from a probability distribution $P(J_{ij})$. The average over disorder is taken on the free energy,
\EQ
\overline{F}=\sum_{\{J_{ij}\}}P(J_{ij})F(J_{ij})\,.
\EN
The well known replica method exploits the fact that $F$ is related to the partition function $Z=\sum_{\{s_i\}}e^{-{\cal H}/T}$ as  $F=-\ln Z$, so that the identity
\EQ
\overline{F}=-\overline{\ln Z}=-\lim_{m\to 0}\frac{\overline{Z^m}-1}{m}
\EN
maps the problem onto that of $m\to 0$ replicas coupled by the average over disorder. Figure \ref{phd} qualitatively shows the phase diagram yielded by numerical simulations (see e.g. \cite{PHP,HPtPV}) for the two-dimensional Ising model with disorder distribution $P(J_{ij})=p\delta(J_{ij}-1)+(1-p)\delta(J_{ij}+1)$.

When approaching a fixed point of the renormalization group the correlation length diverges and the universal properties of the system can be studied directly in the continuum, within the field theoretical framework. For the case we consiser, in which homogeneity of the system is restored by the disorder average, the field theory in question is rotationally invariant, and corresponds to the analytic continuation to imaginary time of a relativistically invariant quantum field theory. We study these field theories within their basis of particle excitations, relying only on symmetry and restricting our attention to fixed points. 

\begin{figure}
\begin{center}
\includegraphics[width=10cm]{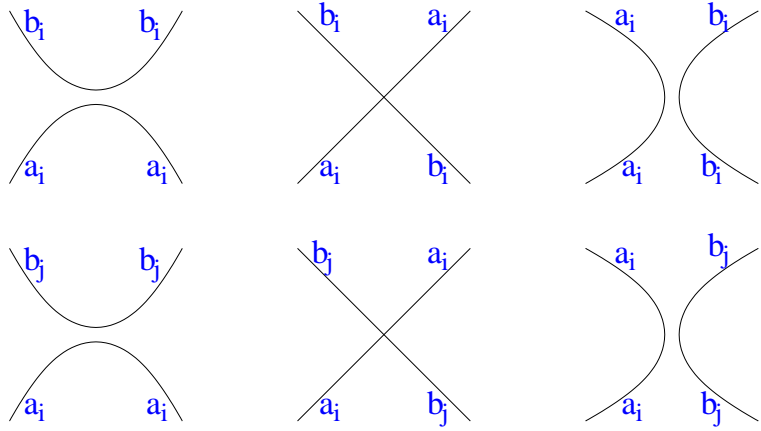}
\caption{Scattering processes corresponding to the amplitudes $S_1$, $S_2$, $S_3$, $S_4$, $S_5$, $S_6$, in that order. Time runs upwards, indices $i$ and $j$ correspond to different replicas. 
}
\label{ampl}
\end{center} 
\end{figure}

As observed in \cite{Zamo_SAW} for the off-critical pure case, $O(N)$ symmetry is implemented adopting a vector multiplet representation of the particle excitations. In our scale-invariant case, these particles are left- and right-movers with momentum and energy related as $p=\pm E$. Moreover, such excitations exist in each of the $m$ replicas and will be denoted as ${a_i}$, where $a=1,2,\ldots N$, $i=1,2,\ldots,m$. When considering the scattering of a right-mover with a left-mover, the infinitely many conservation laws implied by conformal symmetry in two dimensions allow only for final states with a left-mover and a right-mover  \cite{paraf}. The scattering amplitudes are energy independent by scale invariance, and the product of two vectorial representations yields the six possibilities depicted in Fig.~\ref{ampl}. They correspond to transmission and reflection within the same replica ($S_2$ and $S_3$, respectively) or in different replicas ($S_5$ and $S_6$); two identical particles can also annihilate producing another pair within the same replica ($S_1$) or in a different replica ($S_4$). Crossing symmetry \cite{ELOP} then relates amplitudes under exchange of space and time directions as
\bea
S_1=S_3^{*} &\equiv & \rho_{1}\,e^{i\phi}, \\
S_2 = S_2^* &\equiv & \rho_2,\\
S_4 = S_6^* &\equiv & \rho_4\, e^{i\theta}, \\
S_5 = S_5^*&\equiv & \rho_5,
\eea 
where we introduced parametrizations in terms of $\rho_1$ and $\rho_4$ non-negative, and $\rho_2$, $\rho_5$, $\phi$ and $\theta$ real. Finally, unitarity of the scattering matrix translates into the equations
\bea
&& \rho_1^2 + \rho_2^2=1\,, \label{u1}\\
&& \rho_1 \rho_2 \cos\phi=0\,, \label{u2}  \\
&& N \rho_1^2 + N(m-1)\rho_4^2 + 2\rho_1\rho_2 \cos\phi +2\rho_1^2\cos2\phi=0\,,\label{u3} \\
&& \rho_4^2 + \rho_5^{2}=1\,, \label{u4}\\
&& \rho_4 \rho_5 \cos\theta=0\,, \label{u5}\\
&& 2 N \rho_1 \rho_4 \cos(\phi-\theta) + N(m-2)\rho_4^2 + 2\rho_2\rho_4\cos\theta + 2\rho_1\rho_4\cos(\phi+\theta)=0\, \label{u6}. 
\eea

We notice that the superposition $\sum_{a,i}a_ia_i$ scatters into itself with amplitude
\EQ
S=NS_1+S_2+S_3+(m-1)NS_4\,,
\label{singlet}
\EN
which must be a phase by unitarity. Similarly, the combinations $a_ib_i+b_ia_i$ and $a_ib_j+b_ja_i$ scatter into themselves with phases
\bea
\Sigma &=& S_2+S_3\,,
\label{Sigma}\\
\bar{\Sigma} &=& S_5+S_6\,,
\eea
respectively.

The solutions of equations (\ref{u1})--(\ref{u6}) correspond to renormalization group fixed points characterized by $O(N)$ invariance and permutational symmetry of the $m$ replicas. Equations (\ref{u1}) and (\ref{u5}) can be used to express $\rho_2$ and $\rho_5$ in terms of $\rho_1$ and $\rho_4$, which take values in the interval $[0,1]$. The parameter $\rho_4$, to which we refer as disorder modulus, gives a meausre of the disorder strength at the fixed point, since for $\rho_4=0$ the replicas decouple ($S_4=S_6=0$, $S_5=\pm 1$) and Eqs.~(\ref{u1})--(\ref{u3}) are those for the pure case ($m=1$). The interacting solutions for this pure case are \cite{paraf}
\EQ
\rho_1=1, \quad \rho_2=0, \quad -2\cos 2\phi=N\in[-2,2],
\label{pure}
\EN
and
\EQ
\rho_1=\sqrt{1-\rho_2^2}, \quad \cos\phi=0,\quad N=2\,;
\label{bkt}
\EN
the latter is a line of fixed points parametrized by $\rho_1$ that accounts for the Berezinskii-Kosterlitz-Thouless (BKT) phase of the XY model \cite{BKT}. 

\begin{figure}
\begin{center}
\includegraphics[width=7cm]{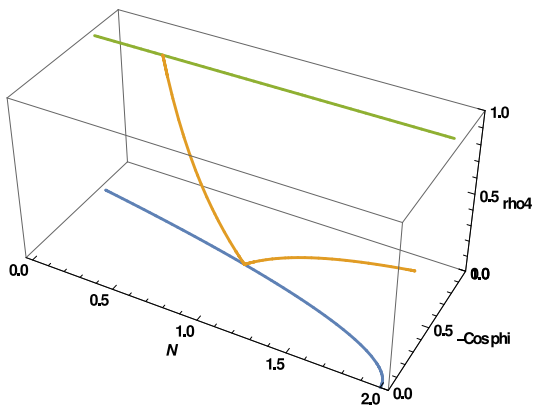}
\caption{Projection in the parameter subspace $\rho_4$-$\cos\phi$ of the lines of renormalization group fixed points of the disordered $O(N)$ model, for $N\in(0,2)$. The disorder modulus $\rho_4$ varies from 0 (pure case) to the maximal value 1. Merging occurs at $N=1$ for $\rho_4=0$ and $N_*=\sqrt{2}-1$ for $\rho_4=1$.
}
\label{space}
\end{center} 
\end{figure}

Coming to random fixed points ($\rho_4\neq 0$, $m=0$), Eq.~(\ref{u3}) shows that they have $\rho_1\neq 0$, and (\ref{u4}), (\ref{u5}) show that they fall into two classes. The first class has $\cos\theta=0$ and disorder modulus varying with $N$, while the second class has fixed (actually maximal) disorder modulus $\rho_4=1$. Considering the class with varying $\rho_4$, we look for the line of fixed points that approaches the pure Ising point as $N\to 1$. Then (\ref{pure}) excludes $\cos\phi=0$ for any $N$, so that (\ref{u2}) implies $\rho_2=0$, and we finally obtain
\EQ
\rho_1=1,\quad \rho_2=\cos\theta=0,\quad \cos\phi=-\frac{1}{N+1},\quad \rho_4=\left|\frac{N-1}{N+1}\right|\sqrt{\frac{N+2}{N}}\,.
\label{varying}
\EN
For positive $N$ this solution is defined for $N\geq\sqrt{2}-1$, and has $\rho_4\to 0$ as $N\to 1$, as expected. Notice that for this solution the phase (\ref{singlet}) becomes $S=2\cos\phi=-1$ at $N=1$, in agreement with the fact that the pure Ising model in two dimensions is a free fermionic theory (scattering on the line involves position exchange); actually, this has been used to fix the sign of $\cos\phi$ in (\ref{varying}). We know from Harris criterion\footnote{The scaling dimension $X_\varepsilon$ of the energy density operator in the pure model becomes smaller than 1 for $N<1$ (see e.g. \cite{paraf}), so that weak disorder, with scaling dimension $2X_\varepsilon$, is relevant.} that the branch with $N<1$ is a line of infrared fixed points, and we see that it extends down to the minimal value $N_*=\sqrt{2}-1$. At this point the solution (\ref{varying}) has $\rho_4=1$ and reaches the subspace of fixed points with maximal disorder modulus (Fig.~\ref{space}). In this subspace there exists and is unique a solution coinciding with (\ref{varying}) at $N_*$; it is defined for any $N$ and reads
\EQ
\rho_1=\rho_4=1,\quad \rho_2=0,\quad \cos\phi=-\frac{1}{\sqrt{2}},\quad \cos\theta=-\frac{N^2+2N-1}{\sqrt{2}(N^2+1)}\,.
\label{strong}
\EN
The subspace with $\rho_4=1$ contains another solution defined for positive $N$, and actually for any $N$; it differs from (\ref{strong}) for having $\cos\theta=\cos\phi$, and is then completely $N$-independent.

The fixed point pattern of Fig.~\ref{space} allows a discussion of the renormalization group flows between $\rho_4=0$ and $\rho_4=1$. First of all we know that weak disorder is relevant for $N\in(0,1)$ and irrelevant for $N\in(1,2)$. This means that for $N\in(0,N_*)$ the flow goes directly from the pure model to the strong disorder solution (\ref{strong}), while for $N\in(N_*,1)$ there are flows from the pure model and the solution (\ref{strong}) towards the infrared fixed line (\ref{varying}). For $N\in(1,2)$, on the other hand, there are flows towards the pure model both from the solution (\ref{varying}) and from one of the solutions with $\rho_4=1$, and we expect this pattern to extend to the region\footnote{For $N>N_*$ the solution (\ref{varying}) approaches $\rho_4=1$ only in the asymptotic limit $N\to\infty$.} $N>2$. Indeed, for $N>2$ the unitarity equations for the pure model admit only the free solutions $S_1=S_3=0$, $S_2=\pm 1$, consistently with the fact that the pure model with $N>2$ only possesses an asymptotically free zero-temperature fixed point, which exhibits an exponentially diverging correlation length (see e.g. \cite{Cardy_book}). The latter property means that the energy density operator of the pure model is marginal, so that weak disorder is irrelevant. 

To these flows we have to add those at $\rho_4=1$ between solutions differing for the value of the second disorder parameter $\theta$. Taking this into account, for values of $N$ inside the interval $(N_*,1)$ the theory naturally accounts for a pattern of three flows between four fixed points (critical point of the pure model, infrared fixed point, multicritical point, zero temperature infrared fixed point) as that observed numerically at $N=0.6$ in \cite{SJK}. The role of $\theta$ will be discussed in more detail in \cite{DL2}, where we will also give the solutions of the fixed point equations for finite number of replicas, and will discuss the case $N=0$, relevant for polymers in a disordered environment. 

It is interesting to notice that in \cite{SJK} the phase diagram was also numerically explored for $N=8$, with results that might appear not completely consistent with what we found for the regime $N>2$. The point can be illustrated for the pure case, where only a fixed point with $\mathbb{Z}_3$ symmetry was observed in \cite{SJK}, while we saw that no such a fixed point is allowed by $O(N)$ symmetry. The explanation is in the fact that the study of \cite{SJK} is made for the loop model on the hexagonal lattice. It is well known that the partition function of the $N$-vector spin model can be rewritten as a sum over loop configurations \cite{DeGennes,Cardy_book}. If this is done on the hexagonal lattice \cite{Nienhuis}, the loops cannot intersect. As originally observed in \cite{Zamo_SAW}, the loop paths correspond in the scattering picture to the particle trajectories, and non-intersection in the pure model corresponds to $S_2=0$ (see Fig.~\ref{ampl}). It follows that the hexagonal lattice loop model yields the fixed points of the pure $O(N)$ spin model in the interval $N\in(-2,2)$, where $S_2=0$ (see Eq.~(\ref{pure})), but not in the regime $N>2$, where there is no reflection at all. For $N>2$ the hexagonal lattice loop model only exhibits a $\mathbb{Z}_3$-symmetric fixed point associated to the specific lattice symmetry rathen than to $O(N)$ symmetry \cite{GBW,SJK}. 

For $N=2$ the equations (\ref{u1})--(\ref{u6}) admit a line of fixed points only in the pure case $\rho_4=0$; this is the line (\ref{bkt}) that, as we already pointed out, accounts for the BKT phase of the pure model. On the other hand, since the flow from $\rho_4=1$ can end in the infrared onto any point of the line (\ref{bkt}), also the disordered model should exhibit a BKT phase, and this is confirmed by numerical studies (see e.g. \cite{APV,OYSE}). The phase diagram observed in these studies is similar to that of Fig.~\ref{phd}, with the ferromagnetic phase replaced by the BKT phase\footnote{Notice that the model studied in \cite{APV,OYSE} is the random phase XY model, for which ${\bf s}_i=(\cos\alpha_i,\sin\alpha_i)$, the nearest neighbor interaction is $-\cos(\alpha_i-\alpha_j+A_{ij})$, and $A_{ij}$ are the random variables drawn from a distribution $P(A_{ij})\propto e^{-A^2_{ij}/\sigma}$; $\sigma$ replaces $1-p$ in Fig.~\ref{phd}. Relying only on symmetry, our formalism applies also to this type of disorder.}. On the other hand, numerical studies still disagree on the values of critical exponents along the portion of the phase boundary going from the multicritical point $M$ to the critical point of the pure model: a constant magnetic exponent $\eta=1/4$ (the value at the BKT transition in the pure model) was deduced in \cite{APV}, while a continuously varying $\eta$ was found in \cite{OYSE}. 

It can be checked that the scattering phase (\ref{singlet}) is $N$-independent for the solution (\ref{varying}), and that $N$ dependence disappears only in the limit $m=0$ corresponding to quenched disorder. This means that the symmetry sector of the superposition $\sum_{a,i}a_ia_i$, to which the energy density operator belongs, becomes superuniversal along this line of fixed points. An analogous result obtained in \cite{random} and further discussed in \cite{potts_random} accounts for the accumulated evidence  \cite{CFL,DW,KSSD,CJ,CB2,OY,JP,Jacobsen_multiscaling,AdAI} that the correlation length critical exponent $\nu$ in the random bond $q$-state Potts ferromagnet does not show any appreciable deviation from the Ising value up to $q$ infinite. On the other hand, the spin operator does not belong to the superuniversal sector and its scaling dimension is expected to vary along the solution (\ref{varying}). This scaling dimension was measured in \cite{SJK} at $N=0.55$ on the infrared fixed line and found to be consistent with the two-loop perturbative result of \cite{Shimada}. We also observe that the phase amplitude (\ref{Sigma}) is straightforwardly seen to be $N$-independent along the solution (\ref{strong}). 

In summary, we used scale (as well as conformally) invariant scattering theory to exactly determine the lines of renormalization group fixed points in $O(N)$ invariant models with quenched disorder. We showed that random fixed points are characterized, in particular, by two disorder parameters: a modulus $\rho_4$ and a phase angle $\theta$. The critical lines fall into the three classes with $\rho_4=0$ (pure case), $\rho_4=1$ (containing Nishimori-like multicritical points as well as zero temperature fixed points), and $\rho_4$ interpolating between 0 and 1 as $N$ varies. The pattern of fixed points allowed us to deduce, in particular, that weak disorder drives the system to $\rho_4=1$ for values of $N$ in the interval $(0,N_*=\sqrt{2}-1)$, to a line of fixed points of interpolating type in the interval $(N_*,1)$, and to the pure system for $N>1$. The exact result $N_*=0.414..$ is not far from the numerical estimate $N_*\approx 0.5$ obtained in \cite{SJK}. The infrared line spanning the interval $(N_*,1)$ exhibits superuniversality in the symmetry sector of the energy density operator. For $N=2$ a line of fixed points exists only in the pure system and accounts also for the BKT phase observed in the random case.


\end{document}